 \documentclass[final,5p,times,twocolumn]{elsarticle}

 \usepackage{graphicx}

\usepackage{amssymb}

\biboptions{square,sort&compress}

\journal{Astroparticle Physics}

\begin{document}

\begin{frontmatter}



\title{Power Spectrum Analysis of LMSU (Lomonosov Moscow State University) Nuclear Decay-Rate Data: Further Indication of r-Mode Oscillations in an Inner Solar Tachocline}

 \author[pas1]{Peter A. Sturrock\corref{cor1}}
 \ead{sturrock@stanford.edu}
\cortext[cor1]{Corresponding author; Tel +1 6507231438; fax +1 6507234840} 
 \address[pas1]{Center for Space Science and Astrophysics, Stanford University, Stanford, CA 94305-4060, USA}

\author[agp1]{Alexander G. Parkhomov}
 \ead{alexparh@mail.ru}
 \address[agp1]{Institute for Time Nature Explorations, Lomonosov Moscow State University, Moscow, Russia}
 
 \author[ef1]{Ephraim Fischbach}
 \ead{ephraim@purdue.edu}
 \address[ef1]{Department of Physics, Purdue University, West Lafayette, IN 47907, USA}
 
 \author[jhj1]{Jere H. Jenkins}
 \ead{jere@purdue.edu}
 \address[jhj1]{School of Nuclear Engineering, Purdue University, West Lafayette, IN 47907}

%

\begin{abstract}
This article presents a power-spectrum analysis of 2,350 measurements of the $^{90}$Sr/$^{90}$Y decay process acquired over the interval 4 August 2002 to 6 February 2009 at the Lomonosov Moscow State University (LMSU). As we have found for other long sequences of decay measurements, the power spectrum is dominated by a very strong annual oscillation. However, we also find a set of low-frequency peaks, ranging from 0.26 year$^{-1}$ to 3.98 year$^{-1}$, which are very similar to an array of peaks in a power spectrum formed from Mt Wilson solar diameter measurements. The Mt Wilson measurements have been interpreted in terms of r-mode oscillations in a region where the sidereal rotation frequency is 12.08 year$^{-1}$. We find that the LMSU measurements may also be attributed to the same type of r-mode oscillations in a solar region with the same sidereal rotation frequency. We propose that these oscillations occur in an inner tachocline that separates the radiative zone from a more slowly rotating solar core.
\end{abstract}


\end{frontmatter}


\section{Introduction}
\label{Intro}

Analyses of decay rates of radioactive elements acquired at Time Nature Explorations at the Lomonosov Moscow State University (LMSU) have revealed evidence for variability, specifically oscillations with periods of one year and of approximately one month \cite{par04}. We have noted \cite{jen09} that an annual periodicity is also exhibited by data acquired at the Brookhaven National Laboratory (BNL) \citep{alb86} and by data acquired at the Physikalisch-Technische Bundesanstalt (PTB) \cite{sie98}. Power spectrum analysis of the BNL and PTB datasets also reveals a modulation with a period of order one month, which may be due to solar rotation \cite{stu10a,jav10,stu10b}. It is also probable that the modifications of decay rates of radioactive nuclides are influenced by the relic neutrino flux \cite{par10}.

This article presents an analysis of an approximately daily compilation of 2,350 measurements of the $^{90}$Sr/$^{90}$Y decay process acquired at the Institute for Time Nature Explorations at the Lomonosov Moscow State University over the time interval August 4 2002 (2002.59) to February 6 2009 (2009.10) \cite{par11}. In order to obtain a smoothly running set of dates that are close to calendar dates, we count days from 1970 January 1 as day 1 and then convert to years by dividing by 365.2564 and adding 1970. (This procedure has proved convenient for the analysis of neutrino data, which all fall into this timeframe.)

The possibility that the solar core may rotate more slowly than the radiative zone (that has a sidereal rotation rate of about 13.5 year$^{-1}$) has led us to conjecture that there may be an ``inner tachocline,'' separating the core from the radiative zone \cite{stu11}. By analogy with the ``outer 
tachocline'' that separates the radiative zone from the convection zone, this would be a localized region where there is a sharp radial gradient in the rotation rate. 

The well known Rieger oscillation \cite{rie84} may be interpreted as an r-mode oscillation \cite{sai82} in the outer tachocline \cite{stu99}. These oscillations are retrograde waves in a rotating fluid with frequencies determined by the sidereal rotation frequency, $\nu_R$, and the values of $l$ and $m$, two of the three spherical harmonic indices. The values of $l$ and $m$ are restricted by $l=2,3,\ldots,m=1,2,\ldots,l$.  The frequencies relevant to processes within the Sun are given by

\begin{eqnarray}\label{eq:eq1}
\nu\left(l,m\right)=\frac{2m\nu_R}{l\left(l+1\right)}
\end{eqnarray}

\noindent{}The Rieger oscillation, with frequency 2.37 year$^{-1}$ corresponding to a period of 154 days, may be interpreted as an r-mode oscillation with $l=3$, $m=1$, where the sidereal rotation frequency is 14.2 year$^{-1}$. This rotation frequency identifies the source of these oscillations as the outer tachocline \cite{sch02}.

	We have recently carried out a power-spectrum analysis of a sequence of 39,024 measurements of the solar diameter made at the Mount Wilson Solar Observatory \cite{stu10c}. This analysis yields strong evidence for an array of oscillations that may be attributed to a set of  r-mode frequencies (for $m=1, l=2,3,4,5,6,7,8,10$) originating in an internal location where the sidereal rotation frequency is 12.08 year$^{-1}$. This estimate also is suggestive of a source in the inner tachocline.

The present article is concerned with measurements of the $^{90}$Sr/$^{90}$Y decay process acquired by Time Nature Explorations at the Lomonosov Moscow State University (LMSU). Parkhomov has found that these measurements reveal evidence for two oscillations, one with a period of one year and the other with a period of approximately one month, and suggests that the annual modulation may be due in part to a cosmic-neutrino component of dark matter \cite{par10,par11}.

We carry out a power-spectrum analysis of the LMSU dataset in Section \ref{sec:PSA}. We search the power spectrum for evidence of r-mode oscillations in Section \ref{sec:rmode}, and we discuss the results in Section \ref{sec:Disc}.

\section{Power Spectrum Analysis}
\label{sec:PSA}

\begin{figure}[h]
 \includegraphics[width=\columnwidth]{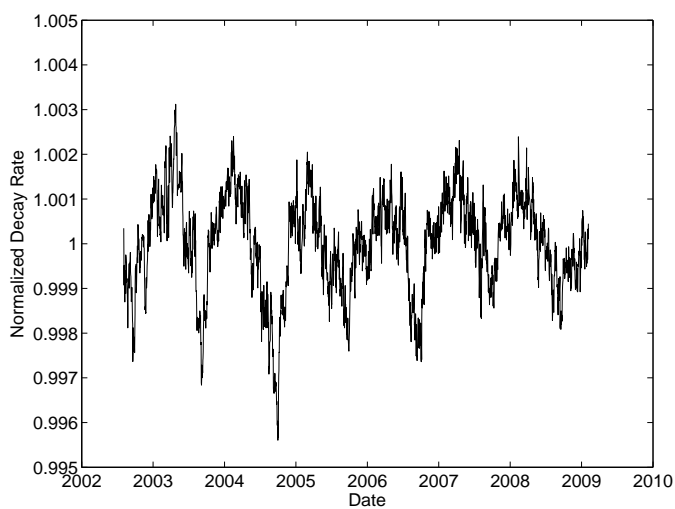}
 \caption{$^{90}$Sr/$^{90}$Y decay measurements, after adjusting for mean decay rate and normalizing.}
 \label{fig:fig1}
\end{figure}

We have determined, and corrected for, the mean decay rate of the LMSU measurements of the $^{90}$Sr/$^{90}$Y decay process, and then normalized the corrected measurements to mean value unity. The resulting normalized measurements are shown as a function of time in Fig. \ref{fig:fig1}. The standard deviation of these data is 0.11\%.

\begin{figure}[h]
 \includegraphics[width=\columnwidth]{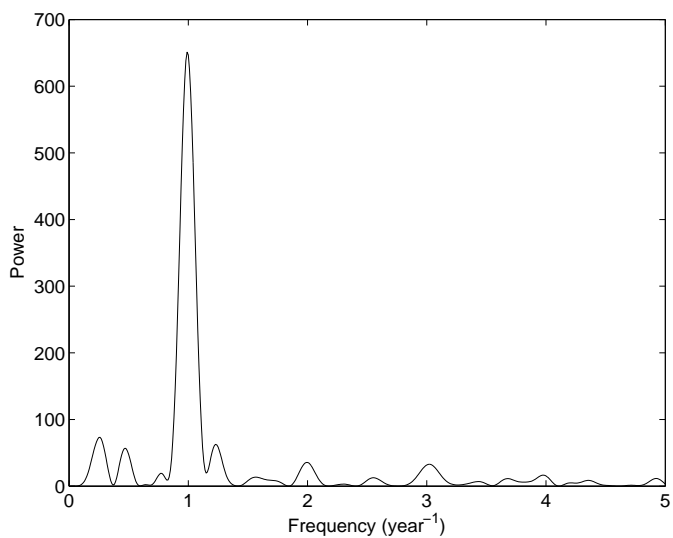}
 \caption{Power spectrum analysis of the normalized data. The peak at 1 year$^{-1}$ has power 651.}
 \label{fig:fig2}
\end{figure}

We next carried out a power-spectrum analysis of these normalized measurements by a likelihood procedure \cite{stu05a}  that is an extension of the Lomb-Scargle procedure \cite{lom76,sca82}. The resulting power spectrum, restricted to the frequency range 0-5 year$^{-1}$, is shown in Fig. \ref{fig:fig2}. The power spectrum is dominated by a very strong peak at 1 year$^{-1}$, as is typical of decay-rate measurements \cite{stu10a,jav10,stu10b}.

\begin{figure}[h]
 \includegraphics[width=\columnwidth]{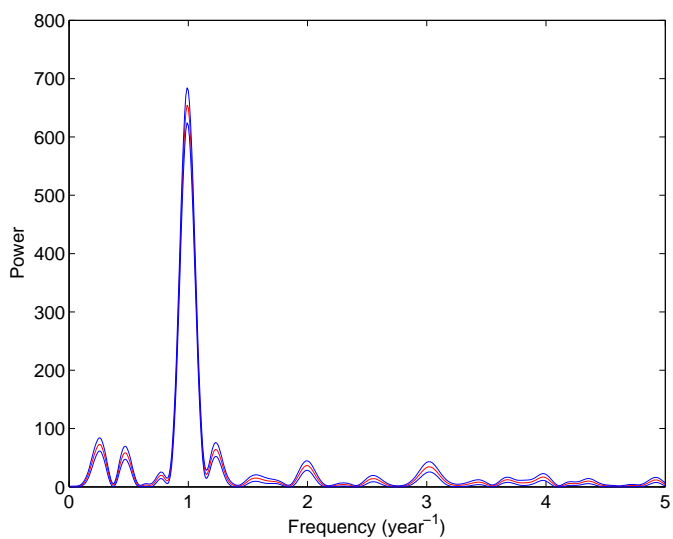}
 \caption{Power spectrum formed from 100 bootstrap simulations of the normalized data. Mean value in red, mean $\pm$ standard deviation in blue.}
 \label{fig:fig3}
\end{figure}

If the dataset comprised both measurements and error estimates, we could examine the robustness of the power spectrum by sampling the data many times, drawing each measurement from a normal distribution with the appropriate mean and standard deviation. We do not have error estimates for the dataset we are examining, but we may examine the robustness of the power spectrum by using the bootstrap procedure \cite{efr93}. Given $N$ time-measurement pairs, we may extract one pair randomly, then draw another pair randomly, having replaced the first pair, and so on, until one has constructed a simulation of the $N$ time-measurement pairs. We then form the power spectrum of this simulated dataset. We have carried out 100 such bootstrap simulations, from which we form the mean power and the standard deviation of the power at each frequency. The result is shown in Fig. \ref{fig:fig3}, from which it is clear that the uncertainty in the power spectrum, estimated in this way, is only a minor factor in our analysis. We therefore ignore this uncertainty from now on.

\section{Evidence for r-Mode Oscillations}
\label{sec:rmode}

We now seek to determine whether there is any value of the rotation frequency $\nu_R$ for which the power spectrum shows peaks corresponding to values of the r-mode frequencies given by Eq. \ref{eq:eq1}. This requires the simultaneous examination of the power at several different frequencies. A convenient procedure is to use the ``Combined Power Statistic'' \cite{stu05b}, defined as follows: If $U$ is the sum of $n$ powers,

\begin{eqnarray}\label{eq:eq2}
U=S_1+S_2+\ldots+S_n,
\end{eqnarray}

\noindent{}then the CPS is defined by

\begin{eqnarray}\label{eq:eq3}
G\left(U\right)=U-\ln \left(1+U+\frac{1}{2}U^2+\ldots + \frac{1}{\left(n-1\right)!}U^{n-1}\right).
\end{eqnarray}

\noindent{}This statistic has the following property: If each power is drawn from an exponential distribution, as is appropriate if the power is formed from a time series dominated by normally distributed random noise \cite{sca82}18], then $G$ also conforms to an exponential distribution. 

\begin{figure}[h]
 \includegraphics[width=\columnwidth]{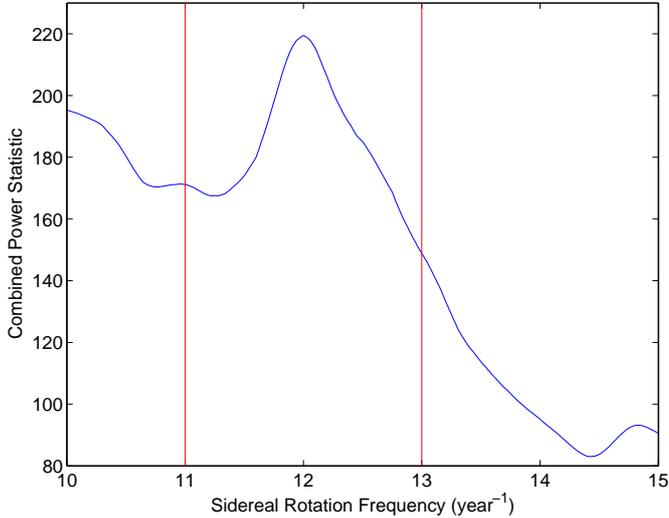}
 \caption{The Combined Power Statistic, formed form $m=1, l=2,3,4,5,6,7,8,9$, r-mode frequencies. The peak is found at $\nu$=12.08 year$^{-1}$, with CPS value 215.6. }
 \label{fig:fig4}
\end{figure}

We show this statistic in Fig \ref{fig:fig4} for $\nu_R$ in the range 10-15 year$^{-1}$ (which is the search band we have adopted, as a convention, for internal solar rotation rates), combining power measurements for the following r-modes: $m=1, l=2,3,4,5,6,7,8,9$. We see that there is a well-defined peak close to $\nu_R$=12.0 year$^{-1}$.

\begin{figure}[h]
 \includegraphics[width=\columnwidth]{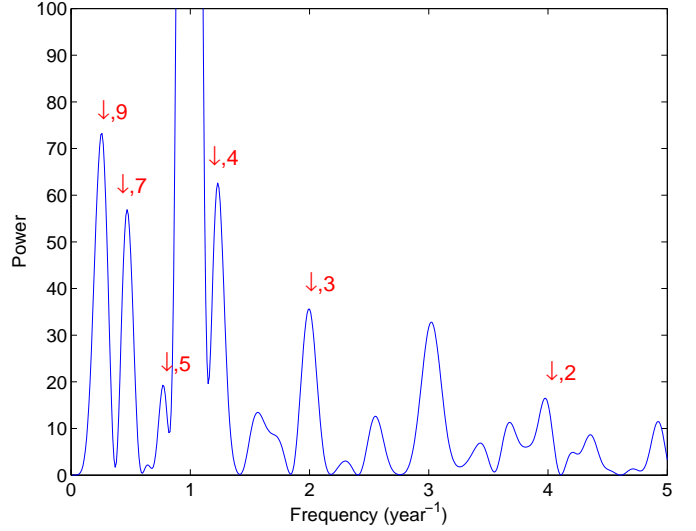}
 \caption{The Combined Power Statistic, formed form $m=1,l=2,3,4,5,6,7,8,9$, r-mode frequencies. The peak is found at $\nu$=12.08 year$^{-1}$, with CPS value 219.5. }
 \label{fig:fig5}
\end{figure}

Figure \ref{fig:fig5} shows the power spectrum over the frequency range 0-5 year$^{-1}$, with arrows indicating the r-mode frequencies for $m=1,l=2,3,4,5,7,9$, and for $\nu_R$=12.08 year$^{-1}$, which is the sidereal rotation rate derived from our analysis of Mt Wilson diameter data \cite{stu10c}. We see that there is good agreement between these values and peaks in the power spectrum. (There are no peaks corresponding to $m=1, l=6~\rm{and}~8$.)

\begin{figure}[h]
 \includegraphics[width=\columnwidth]{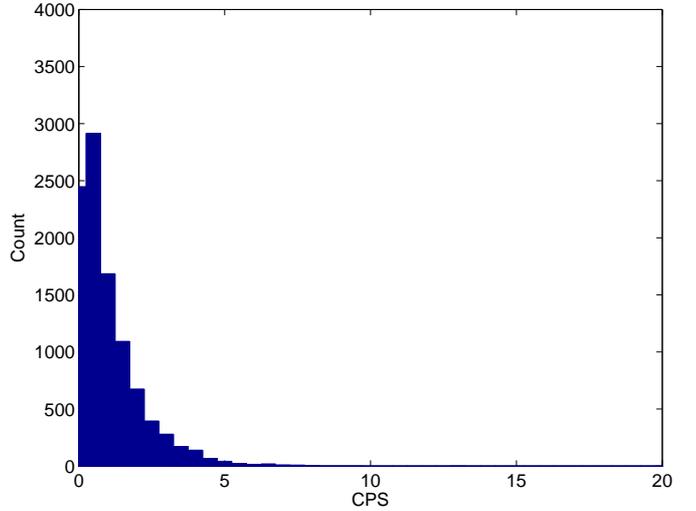}
 \caption{Histogram of the combined power statistic formed from 10,000 shuffle-simulations for $\nu$=12.08 year$^{-1}$. None is larger than 12.8, whereas the actual value is 215.6.}
 \label{fig:fig6}
\end{figure}

As a test of the significance of this result, we have carried out 10,000 shuffle simulations \citep{bah91} of the data, with the result shown in histogram form in Fig. \ref{fig:fig6}. None is larger than 12.8, whereas the actual value is 215.6.

\section{Discussion}
\label{sec:Disc}

\begin{figure}[h]
 \includegraphics[width=\columnwidth]{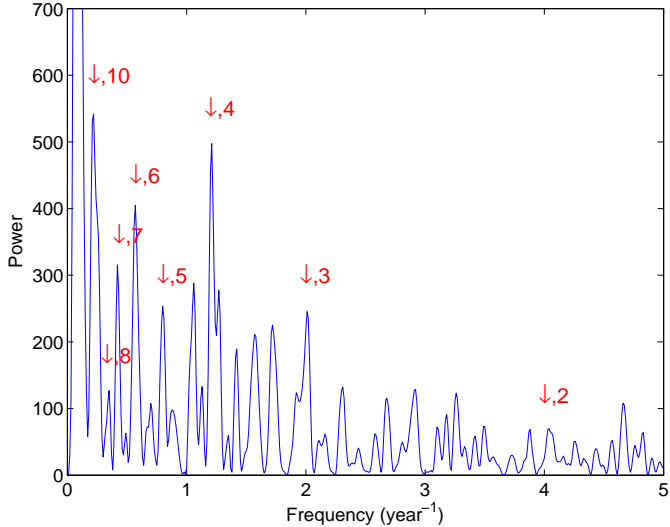}
 \caption{Power spectrum formed from Mt Wilson diameter data, showing the locations of peaks expected for r-modes for sidereal rotation frequency 12.08 year$^{-1}$. Red arrows indicate $m=1,l=2,3,4,5,7,8,10$.}
 \label{fig:fig7}
\end{figure}

We show in Fig. \ref{fig:fig7} the power spectrum recently derived from Mt Wilson diameter measurements \cite{stu10c}. This is remarkably similar to the power spectrum derived from $^{90}$Sr/$^{90}$Y data, shown in Fig. \ref{fig:fig5}, suggesting that they have a common origin. Each of these datasets exhibits periodicities that may be attributed to a complex of r-mode oscillations. The Mt Wilson data show evidence of oscillations corresponding to $m=1,l=2,3,4,5,6,7,8,10$ and a sidereal rotation frequency of 12.08 year$^{-1}$, the peak at $l=9$ being missing. By contrast, we see from Fig. \ref{fig:fig5} that, in the LMSU power spectrum, the peaks at $l=6~\rm{and}~8$ are missing. These missing modes may not be significant, since studies of the Rieger \cite{rie84} and similar oscillations show that solar r-modes oscillations are quite typically intermittent.

As mentioned in the Introduction, the Rieger and related oscillations may be attributed to r-mode oscillations that develop in a region where the sidereal rotation rate is near 13.7 year$^{-1}$, placing it in the familiar tachocline \cite{sch02}. This suggests that r-mode oscillations are unstable and can grow to large amplitudes in regions where there is a radial gradient in angular velocity. (In this connection, it is interesting to note that Zaqarashvili et al. have shown theoretically that a \textit{latitudinal} gradient in angular velocity can lead to instability \cite{zaq10}.) Since the r-mode oscillations manifested in the LMSU and Mt Wilson data seem to occur where the sidereal rotation rate is close to 12 year$^{-1}$ (much less than the rotation rate of the radiative zone), these results are compatible with the proposal that these oscillations develop in an \textit{inner tachocline} that separates the core from the radiative zone.

\begin{figure}[h]
 \includegraphics[width=\columnwidth]{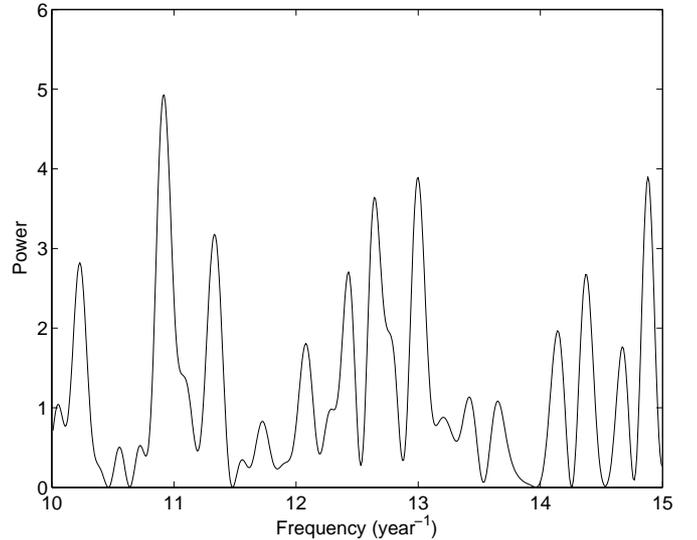}
 \caption{Power spectrum formed from LMSU data for the frequency range 10-15 year$^{-1}$. The biggest peak is found at 10.92 year$^{-1}$, with power $S$=4.93.}
 \label{fig:fig8}
\end{figure}

We have recently presented the results of power-spectrum analyses of nuclear decay data acquired at the Brookhaven National Laboratory (BNL) \cite{stu10a,jav10} and at the Physikalisch-Technische Bundesanstalt laboratory (PTB) \cite{jav10,stu10b}. Analysis of BNL data yields evidence of periodicities at or near 11.2 year$^{-1}$ and 11.9 year$^{-1}$ \cite{stu10a}. When the BNL analysis is combined with an analysis of PTB data, we find the dominant oscillation to be at about 11.2 year$^{-1}$ \cite{stu10b}. If this is interpreted as a rotational modulation, we see that these oscillations originate in a region with a sidereal rotation rate of about 12.2 year$^{-1}$, not far from the estimates of rotation rates inferred from our analysis of r-mode oscillations. Fig. \ref{fig:fig8} shows the power spectrum formed from LMSU data for the frequency range 10-15 year$^{-1}$ We see that the biggest peak in this range is one at 10.92 year$^{-1}$ with power $S = 4.93$. If interpreted as a rotational modulation, it corresponds to a sidereal rotation frequency of 11.92 year$^{-1}$, not far from the value (12 year$^{-1}$) inferred from our r-mode analysis.

Concerning internal rotation rates, it is interesting to review the results of power spectrum analyses of the Super-Kamiokande solar neutrino data \cite{fuk03}. The most prominent peak in a power spectrum analysis of Super-Kamiokande data (in 5-day bins) is found at 9.43 year$^{-1}$ \cite{stu05b}. If this peak is attributed to rotational modulation, it indicates that the Sun contains a region with a sidereal rotation rate of 10.43 year$^{-1}$, which is suggestive of a slowly rotating core. 

In discussing the power spectrum of Super-Kamiokande data, it is important to note that if these data are analyzed by the Lomb-Scargle procedure (as in the analysis carried out by the Super-Kamiokande Consortium \cite{yoo03}), this peak does not appear to be statistically significant. However, if the data are analyzed by a likelihood procedure that takes account of the error estimates and of the specific durations of the time-bins (which the Lomb-Scargle procedure does not), the peak is found to be quite significant \cite{stu05b}.

If the inner tachocline extends from the core, with a rotation rate of about 10.4 year$^{-1}$, to the radiative zone, with a rotation rate of about 13.7 year$^{-1}$, we see that the rotation rate has the value 12 year$^{-1}$ in the center of the tachocline. There is a correspondence between this result and the fact that the Rieger and related oscillations seem to have their origin in the center of the outer tacholine.

\section*{Acknowledgements}
The work of EF was supported in part by U.S. DOE contract No. DE-AC02-76ER071428.












\end{document}